\documentclass[journal]{IEEEtran}
\usepackage{cite}
\usepackage[dvips]{graphicx}
\DeclareGraphicsExtensions{.eps}
\usepackage{amsmath}
\usepackage{amssymb}
\usepackage{cite}

\newcommand{\diag}{\mathrm{diag}}
\newcommand{\ve}{\mathrm{vec}}
\newcommand{\T}{\mathrm{T}}
\newcommand{\He}{\mathrm{H}}
\newcommand{\jim}{\jmath}

\newenvironment{mat}[1]{\left[\begin{array}{#1}}{\end{array}\right]}
\newcommand{\bmx}[1]{\begin{mat}{#1}}
\newcommand{\emx}{\end{mat}}

\newcommand{\gssa}[4]{\mbox{\boldmath $#1$}_{#2}^{#3}(#4)}

\newcommand{\gss}[3]{\mbox{\boldmath $#1$}_{#2}^{#3}}
\newcommand{\g}[1]{\mbox{\boldmath $#1$}}

\newcommand{\sbm}[1]{\mbox{\scriptsize $\g{#1}$}}

\newcommand{\pref}[1]{(\ref{#1})}

\begin{document}

\title{Joint Channel Estimation / Data Detection in MIMO-FBMC/OQAM Systems -- A Tensor-Based Approach}

\author{\IEEEauthorblockN{Eleftherios Kofidis\IEEEauthorrefmark{1}\IEEEauthorrefmark{2},\thanks{Part of this work was done while the first author was visiting the Group Science, Engineering and Technology, at KULeuven-Kulak, Belgium.}
Christos Chatzichristos\IEEEauthorrefmark{3}\IEEEauthorrefmark{2} and Andr\'{e} L.\ F.\ de Almeida\IEEEauthorrefmark{4}}
\IEEEauthorblockA{\IEEEauthorrefmark{1}Department of Statistics and Insurance Science,\\
        University of Piraeus, Piraeus, Greece\\
				E-mail: kofidis@unipi.gr\\
				\IEEEauthorrefmark{2}Computer Technology Institute \& Press ``Diophantus" (CTI), Patras, Greece\\
				\IEEEauthorrefmark{3}Department of Informatics and Telecommunications,\\
        University of Athens, Athens, Greece\\
				E-mail: chrichat@hotmail.com\\
				\IEEEauthorrefmark{4}Department of Teleinformatics Engineering,\\
Federal University of Cear\'{a}, Fortaleza, Brazil\\
				E-mail: andre@gtel.ufc.br}}
				
\maketitle

\begin{abstract}
Filter bank-based multicarrier (FBMC) systems are currently being considered as a prevalent candidate for replacing the long established cyclic prefix (CP)-based orthogonal frequency division multiplexing (CP-OFDM) in the physical layer of next generation communications systems. In particular, FBMC/OQAM has received increasing attention due to, among other features, its potential for maximum spectral efficiency. It suffers, however, from an intrinsic self-interference effect, which complicates signal processing tasks at the receiver, including synchronization, channel estimation and equalization. In a multiple-input multiple-output (MIMO) configuration, the multi-antenna interference has also to be taken into account. (Semi-)blind FBMC/OQAM receivers have been little studied so far and mainly for single-antenna systems. The problem of joint channel estimation and data detection in a MIMO-FBMC/OQAM system, given limited or no training information, is studied in this paper through a tensor-based approach in the light of the success of such techniques in OFDM applications. Simulation-based comparisons with CP-OFDM are included, for realistic transmission models. 
\end{abstract}

\section{Introduction}

Filter bank-based multicarrier (FBMC) systems~\cite{f11} are currently being considered as a prevalent candidate for replacing the long established orthogonal frequency division multiplexing (OFDM) in the physical layer of next generation communications systems~\cite{bbcmru14}.  
The potential of FBMC transmission stems from its increased ability to carrying a flexible spectrum shaping, along with a major increase in spectral efficiency and robustness to synchronization requirements, features of fundamental importance in the envisaged networks. 
A particular type of FBMC, known as FBMC/OQAM (or OFDM/OQAM) system, consisting of pulse shaped OFDM carrying offset quadrature amplitude modulation (OQAM) symbols, has received increasing attention due to, among other features, its potential for maximum spectral efficiency~\cite{f11}. Notably, it allows a cyclic prefix (CP)-free transmission while offering very good spectral agility and time localization with very important implications in the system design and performance. It suffers, however, from an \emph{intrinsic} inter-carrier/inter-symbol interference (ICI/ISI), which complicates signal processing tasks at the receiver, including synchronization, channel estimation and equalization~\cite{k14}. Although FBMC/OQAM research has been rapidly advancing in the last decade or so, resulting in a number of well performing techniques for receiver design, (semi-)blind FBMC/OQAM methods have been very little studied so far (e.g., \cite{bdh01,hc14,s15,sbp16}) and mainly for the single-antenna case. Interestingly, (semi-)blind multiple-input multiple-output (MIMO) techniques have been considered as a potential solution to the pilot contamination problem in massive MIMO FBMC-based configurations~\cite{mfmfcf15}. 

Tensor models and methods have been extensively studied for communications applications~\cite{a07}, including system modeling and receiver design of single-input multiple-output (SIMO) and MIMO systems, both in a general~\cite{a07,afm06,rh10,shr13,vsd16} and a multicarrier and/or spread spectrum~\cite{sgb00,bda03,js03,rcs06,nl07,sh10,zgx10,xfd11,fzwlx11,lcas12,xcj12,gl13,lcsa13,sd13,r15,rrc15} setup. The inherent ability of tensor models to capture the relations among the various system's dimensions, in a way that is \emph{unique} under mild conditions and/or constraints, has been exploited in problems of jointly estimating synchronization parameters, channel(s), and transmitted data symbols.  
Tensorial approaches have proven their unique advantages not only in their `natural' applications in (semi-)blind receivers~\cite{sgb00,js03,nl07,sd13,r15,rrc15} but also in the design of training-based high performance receivers for challenging scenarios~\cite{rh10,rkx12,gl13}. Notably, in OFDM applications~\cite{sb01,js03,gl13}, performance close to that with perfect knowledge of the system parameters has been achieved~\cite{js03}. 

In the light of their successful application in OFDM (semi-)~blind estimation problems, tensor-based techniques are considered here in the context of MIMO-FBMC/OQAM systems. The problem of joint channel estimation and data detection, given limited or no training information, is re-visited through a tensorial approach. 
The main difficulties come from the intrinsic interference effect and the lack of a guard interval (CP), which challenge the receiver design even under the commonly made simplifying assumption of channels of low selectivity, also adopted in this paper. Simulations-based comparisons with CP-OFDM are included, for realistic transmission models.  

\section{System Model}

Consider a system based on FBMC/OQAM with $N_{\mathrm{T}}$ transmit and $N_{\mathrm{R}}$ receive antennas. The synthesis filter bank (SFB) output at the transmit (Tx)  antenna $t$ is given by
\begin{equation}
s^{(t)}(l)=\sum_{m=0}^{M-1}\sum_{n}d^{(t)}_{m,n}g_{m,n}(l),
\label{eq:SFB}
\end{equation}
where $(m,n)$ refers to the $m$th subcarrier and the $n$th FBMC symbol, $d^{(t)}_{m,n}$ are \emph{real} OQAM symbols, $M$ is the (even) number of subcarriers, and 
\[
g_{m,n}(l)=g\left(l-n\frac{M}{2}\right)e^{\jim\frac{2\pi}{M}m\left(l-\frac{L_{g}-1}{2}\right)}e^{\jim\varphi_{m,n}},
\] 
with $g$ being the employed prototype filter impulse response (assumed of unit energy) with length $L_{g}$, and $\varphi_{m,n}=(m+n)\frac{\pi}{2}+mn\pi$~\cite{ssl02}. Moreover, usually $L_{g}=KM$, with $K$ being the overlapping factor. Let $\gss{H}{}{(r,t)}=\bmx{cccc} H^{(r,t)}_{0} & H^{(r,t)}_{1} & \cdots & H^{(r,t)}_{M-1}\emx^{\mathrm{T}}$ be the frequency response of the channel from the Tx antenna $t$ to the receive (Rx) antenna $r$, assumed invariant in time. Assume, as usual, that the noise signals at different Rx antennas are zero mean white Gaussian with variance $\sigma^2$ and uncorrelated with each other (i.e., temporally and spatially white noise). Under the common assumption of a (relatively to $M$) low channel delay spread, the analysis filter bank (AFB) output at the Rx antenna $r$ and at the $(p,q)$ frequency-time (FT) point can be written as~\cite{kkrt13}
\begin{equation}
y^{(r)}_{p,q}=\sum_{t=1}^{N_{\mathrm{T}}}H^{(r,t)}_{p}c^{(t)}_{p,q}+w^{(r)}_{p,q},
\label{eq:y=Hc+w} 
\end{equation}
where $w^{(r)}_{p,q}$ denotes the corresponding noise component, known to be also zero mean Gaussian of variance $\sigma^2$ but correlated in both time and frequency and
\begin{equation}
c^{(t)}_{p,q}=d^{(t)}_{p,q}+\jim\sum_{(m,n)\in\Omega_{p,q}}\langle g \rangle_{m,n}^{p,q} d^{(t)}_{m,n}
\label{eq:cpq}
\end{equation}
is the ``virtual" transmitted symbol (or pseudo-symbol) consisting of the corresponding transmitted symbol plus the (imaginary) interference from its \emph{first-order} FT neighborhood $\Omega_{p,q}$. The interference weights $\langle g \rangle$ are known to be symmetric according to the following pattern~\cite{kkrt13}
\begin{equation}
\begin{array}{rrr} 
(-1)^{p}\delta & -\beta & (-1)^{p}\delta \\
-(-1)^{p}\gamma & d_{p,q} & (-1)^{p}\gamma \\
(-1)^{p}\delta & \beta & (-1)^{p}\delta
\end{array}
\label{eq:pattern}
\end{equation}
 with the horizontal and vertical directions denoting time and frequency, respectively, and the constants $\gamma>\beta>\delta>0$ being \emph{a-priori} computable from $g$ (cf.~\cite{kkrt13} for details). 

Let each Tx antenna transmit $N$ FBMC symbols and let $\gss{D}{}{(t)}=[d^{(t)}_{m,n}]\in\mathbb{R}^{M\times N}$ denote the corresponding frame of OQAM data. The corresponding AFB output at the $r$th Rx antenna can then be written as the $M\times N$ matrix
\begin{equation}
\gss{Y}{}{(r)}=[y_{m,n}^{(r)}]=\sum_{t=1}^{N_{\mathrm{T}}}\mathrm{diag}(\gss{H}{}{(r,t)})\gss{C}{}{(t)}+\gss{W}{}{(r)},
\label{eq:Yr}
\end{equation}
where $\gss{C}{}{(t)}=[c^{(t)}_{m,n}]=\bmx{cccc} \gss{c}{0}{(t)} & \gss{c}{1}{(t)} & \cdots & \gss{c}{N-1}{(t)} \emx\in\mathbb{C}^{M\times N}$ collects the virtual symbols for Tx antenna $t$ and $\gss{W}{}{(r)}=[w^{(r)}_{m,n}]$. One can readily see that the intrinsic interference effect as desribed in~\pref{eq:pattern} can be compactly expressed as follows
\begin{equation}
\gss{C}{}{(t)}=\gss{D}{}{(t)}+\jim \left[\beta \g{E}\gss{D}{}{(t)}+\g{S}(-\gamma \gss{D}{}{(t)}\g{\bar{E}}+\delta\g{\bar{Z}}\gss{D}{}{(t)}\g{\tilde{E}})\right],
\label{eq:C}
\end{equation}
where 
\[
\g{S}=\mathrm{diag}(1,-1,1,-1,\ldots,1,-1),
\]
$\g{E}$ is the circulant $M\times M$ matrix
\[
\g{E}=\left[\begin{array}{ccccccc} 0 & 1 & 0 & 0 & \cdots & 0 & -1 \\ -1 & 0 & 1 & 0 & \cdots & 0 & 0 \\ 
0 & -1 & 0 & 1 & \cdots & 0 & 0 \\ \vdots & \vdots & \ddots & \ddots & \cdots & \ddots & \vdots \\
1 & 0 & 0 & \cdots & 0 & -1 & 0 \end{array}\right],
\]
while $\g{\bar{E}}$ and $\g{\tilde{E}}$ are similarly structured Toeplitz $N\times N$ matrices: 
\begin{eqnarray*}
\boldsymbol{\bar{E}} & = & \left[\begin{array}{ccccccc} 0 & 1 & 0 & 0 & \cdots & 0 & 0 \\ -1 & 0 & 1 & 0 & \cdots & 0 & 0 \\ 
0 & -1 & 0 & 1 & \cdots & 0 & 0 \\ \vdots & \vdots & \ddots & \ddots & \cdots & \ddots & \vdots \\
0 & 0 & 0 & \cdots & 0 & -1 & 0 \end{array}\right], \\
\boldsymbol{\tilde{E}} & = & \left[\begin{array}{ccccccc} 0 & 1 & 0 & 0 & \cdots & 0 & 0 \\ 1 & 0 & 1 & 0 & \cdots & 0 & 0 \\ 
0 & 1 & 0 & 1 & \cdots & 0 & 0 \\ \vdots & \vdots & \ddots & \ddots & \cdots & \ddots & \vdots \\
0 & 0 & 0 & \cdots & 0 & 1 & 0 \end{array}\right].
\end{eqnarray*}
Letting 
\[
\g{Z}=\left[\begin{array}{cc} \boldsymbol{0}_{1\times (M-1)} & 1 \\ \boldsymbol{I}_{M-1} & \boldsymbol{0}_{(M-1)\times 1} \end{array}\right]
\]
denote the $M\times M$ matrix of circular downwards shifting, $\g{\bar{Z}}$ can be expressed as 
\[
\g{\bar{Z}}=\g{Z}+\g{Z}^{-1}=\left[\begin{array}{ccccccc} 0 & 1 & 0 & 0 & \cdots & 0 & 1 \\
1 & 0 & 1 & 0 & \cdots & 0 & 0 \\ 0 & 1 & 0 & 1 & \cdots & 0 & 0 \\ \vdots & \ddots & \ddots & \ddots & \cdots & \ddots & \vdots \\
1 & 0 & 0 & 0 & \cdots & 1 & 0 \end{array}\right],
\]
which is also circulant. Note that $\gss{D}{}{(t)}=\Re\{\gss{C}{}{(t)}\}$. 

\noindent
\emph{Remarks.}
\begin{enumerate}
\item As it is common, it is here assumed that the frame is preceded and followed by inactive inter-frame gaps, which can be taken as FBMC symbols of all zeros, thus resulting in negligible interference among frames~\cite{kkrt13}.
\item Although FBMC/OQAM has proved to be more robust than CP-OFDM to imperfect frequency synchronization~\cite{bn10}, incorporating a carrier frequency offset (CFO) into its signal model can be seen to be less straightforward~\cite{mt16}, especially when considering a frequency-domain model as in~\pref{eq:Yr}. Tensor-based methods for CP-OFDM that also estimate CFO (e.g., \cite{js03,sh10,zgx10,fzwlx11}) are based on time-domain processing instead. Future work for FBMC/OQAM may also rely on time-domain processing, following, for instance, \cite{lgss13}. For the sake of simplicity, and in order to concentrate on joint data/channel estimation, perfect synchronization is assumed in the following. \end{enumerate}

\section{Joint Channel Estimation / Data Detection}

Stacking the $N_{\mathrm{R}}$ matrices~\pref{eq:Yr} in the $MN_{\mathrm{R}}\times N$ matrix
\begin{equation}
\gss{Y}{2}{}=\bmx{cccc} (\gss{Y}{}{(1)})^{\mathrm{T}} & (\gss{Y}{}{(2)})^{\mathrm{T}} & \cdots & (\gss{Y}{}{(N_{\mathrm{R}})})^{\mathrm{\mathrm{T}}} \emx^{\T},
\label{eq:Y2}
\end{equation}
one can write (see also~\cite{js03,fzwlx11})
\begin{eqnarray}
\gss{Y}{2}{} & = & \sum_{t=1}^{N_{\mathrm{T}}}\left(\bmx{c} (\gss{H}{}{(1,t)})^{\T} \\ (\gss{H}{}{(2,t)})^{\T} \\ \vdots \\ (\gss{H}{}{(N_{\mathrm{R}},t)})^{\T} \emx \odot \gss{I}{M}{}\right)\gss{C}{}{(t)}+\gss{W}{2} \nonumber \\
& = & (\g{H} \odot \g{\Gamma})\g{C}+\gss{W}{2}{},
\label{eq:Y2eq}
\end{eqnarray}
where $\gss{W}{2}{}$ is similarly defined, $\g{C}=\bmx{cccc} (\gss{C}{}{(1)})^{\mathrm{T}} & (\gss{C}{}{(2)})^{\mathrm{T}} & \cdots & (\gss{C}{}{(N_{\mathrm{\mathrm{T}}})})^{\mathrm{T}} \emx^{\mathrm{T}}$, 
\begin{eqnarray*}
\g{H} & = & \bmx{cccc} \gss{H}{}{(1,1)} & \gss{H}{}{(2,1)} & \cdots & \gss{H}{}{(N_{\mathrm{R}},1)} \\
\gss{H}{}{(1,2)} & \gss{H}{}{(2,2)} & \cdots & \gss{H}{}{(N_{\mathrm{R}},2)} \\
\vdots & \vdots & \ddots & \vdots \\
\gss{H}{}{(1,N_{\mathrm{T}})} & \gss{H}{}{(2,N_{\mathrm{T}})} & \cdots & \gss{H}{}{(N_{\mathrm{R}},N_{\mathrm{T}})}
\emx^{\mathrm{T}}, \\
\g{\Gamma} & = & 
\underbrace{\bmx{cccc} \gss{I}{M}{} & \gss{I}{M}{} & \cdots & \gss{I}{M}{} \emx}_{N_{\mathrm{T}} \mbox{\ times}},
\end{eqnarray*}
and $\odot$ denotes the Khatri-Rao (columnwise Kronecker) product~\cite{sb01}. If $\g{\mathcal{Y}}$ is the $M\times N\times N_{\mathrm{R}}$ (i.e., frequency $\times$ time $\times$ space) tensor of received signals with entries $\g{\mathcal{Y}}_{m,n,r}=y_{m,n}^{(r)}$, then the matrix $\gss{Y}{2}{}$ above results from vertically stacking its $N_{\mathrm{R}}$ frontal slices 
and~\pref{eq:Y2eq} corresponds to its \emph{canonical polyadic decomposition (CPD)} (also known as PARAFAC), of rank $MN_{\mathrm{T}}$~\cite{sb01,sdfhpf16}. The joint estimation problem can then be stated (in the notation of~\cite{kb09}) as 
\begin{equation}
\min_{\sbm{H},\sbm{C}}\|\g{\mathcal{Y}}-[\![\g{\Gamma},\gss{C}{}{\mathrm{T}},\g{H}]\!]\|_{\mathrm{F}},
\label{eq:prob}
\end{equation} 
where $\|\cdot\|_{\mathrm{F}}$ stands for the (tensor) Frobenius norm~\cite{kb09} and the noise color has been ignored for the sake of simplicity. Stacking instead (in a vertical fashion) the lateral slices of $\g{\mathcal{Y}}$ yields the $MN\times N_{\mathrm{R}}$ matrix 
\begin{eqnarray}
\gss{Y}{3}{}\!\!\!\!\!\! & = & \!\!\!\!\!\!\bmx{cccc} \gssa{Y}{}{(1)}{:,1} & \gssa{Y}{}{(2)}{:,1}  & \cdots & \gssa{Y}{}{(N_{\mathrm{R}})}{:,1} \\
\gssa{Y}{}{(1)}{:,2} & \gssa{Y}{}{(2)}{:,2}  & \cdots & \gssa{Y}{}{(N_{\mathrm{R}})}{:,2} \\ \vdots & \vdots & \ddots & \vdots \\
\gssa{Y}{}{(1)}{:,N} & \gssa{Y}{}{(2)}{:,N}  & \cdots & \gssa{Y}{}{(N_{\mathrm{R}})}{:,N}\emx \nonumber \\
& \!\!= & \!\!\!\!\!\!\bmx{cccc} \mathrm{vec}(\gss{Y}{}{(1)})\! & \mathrm{vec}(\gss{Y}{}{(2)})\! & \cdots\! & \mathrm{vec}(\gss{Y}{}{(N_{\mathrm{R}})})\emx,
\label{eq:Y3}
\end{eqnarray}
where Matlab indexing notation has been used. 
In view of~\pref{eq:Yr} and using analogous arguments as previously, one can write~\cite{sb01}
\begin{equation}
\gss{Y}{3}{}=(\gss{C}{}{\mathrm{T}}\odot\g{\Gamma})\gss{H}{}{\T}+\gss{W}{3}{},
\label{eq:Y3eq} 
\end{equation}
with $\gss{W}{3}{}$ constructed as in~\pref{eq:Y3}. 

The \emph{uniqueness} property of the CPD tensor decomposition, which can be trusted to hold under \emph{mild} conditions~\cite{sb01}, can be taken advantage of in the above setup (in a way analogous to that followed in OFDM; see, e.g., \cite{js03} etc.) to \emph{blindly} estimate the channel matrix $\g{H}$ and recover the virtual symbols $\g{C}$ from the tensor of AFB output signals, $\g{\mathcal{Y}}$. However, in a multiple-input ($N_{\mathrm{T}}\geq 2$) system, the matrix $\boldsymbol{\Gamma}$ has collinear columns and hence a k-rank of~1, which implies that the identifiability of the above CPD model is not guaranteed~\cite{ss07}.\footnote{As shown in~\cite{ss07}, k-rank$\geq 2$ for all the factor matrices is a necessary condition for CPD uniqueness.} Of course, this is a \emph{constrained} CPD model~\cite{fa14}, in the sense that one of the factor matrices is known (constrained) to equal $\boldsymbol{\Gamma}$. Hence a number of related identifiability results might be examined as to their applicability here, including those for CPD with one of the factors \emph{a priori} known (e.g., \cite{js04,dd13,sd15}) as well as uniqueness arguments concerning constrained CPD models (e.g., PARALIND~\cite{bhsl09,fa14}). In fact, a quite similar formulation was presented, for MIMO-OFDM, in~\cite{xfd11}, where identifiability was claimed to hold, however based on a proof of questionable validity.\footnote{Involving the inverse of the rank-deficient matrix $\gss{\Gamma}{}{\T}\g{\Gamma}$.} A simple way to see that the CPD in~\pref{eq:prob} does not enjoy uniqueness for $N_{\mathrm{T}}>1$ is the following. 
Stacking the horizontal slices of the tensor $\g{\mathcal{Y}}$ in the $N_{\mathrm{R}}N\times M$ matrix 
\[
\gss{Y}{1}{}=\bmx{cccc} \gss{Y}{}{(1)} & \gss{Y}{}{(2)} & \cdots & \gss{Y}{}{(N_{\mathrm{R}})} \emx^{\T}
\] 
results in the following alternative way of writing the CPD,
\begin{equation}
\gss{Y}{1}{}=(\g{H}\odot\gss{C}{}{\T})\gss{\Gamma}{}{\T}+\gss{W}{1}{},
\label{eq:Y1eq}
\end{equation}
or equivalently
\begin{equation}
\gss{Y}{1}{}=\sum_{t=1}^{N_{\mathrm{T}}}\bmx{c} (\gss{H}{}{(1,t)})^{\T} \\ (\gss{H}{}{(2,t)})^{\T} \\ \vdots \\ (\gss{H}{}{(N_{\mathrm{R}},t)})^{\T} \emx \odot (\gss{C}{}{(t)})^{\T}+\gss{W}{1}{}
\label{eq:Y1eqMIMO}
\end{equation}
Clearly, there is no way to identify $\g{H}$ and $\g{C}$ from the above (unless additional information is made available). 
However, in the SIMO case, \pref{eq:Y1eq} yields
\begin{equation}
\gss{Y}{1}{}=\g{H}\odot\gss{C}{}{\T}+\gss{W}{1}{},
\label{eq:Y1eqSIMO}
\end{equation}
which shows that the channel and (virtual) symbol matrices can be determined (up to scaling ambiguity) through a \emph{Khatri-Rao factorization} of $\gss{Y}{1}{}$ (e.g., \cite{rh10}).\footnote{The result of the factorization could also be used to provide a better initialization of the ALS procedure. See also Proposition~3.1 in~\cite{sd15}, which ensures uniqueness of the CPD when one of its factors ($\boldsymbol{\Gamma}=\gss{I}{M}{}$) is known and of full column rank. Furthermore, applying Proposition~3.2 of~\cite{sd15}, while making the common assumption that $N\gg M\gg N_{\mathrm{R}}$, leads to the trivial requirement that $N_{\mathrm{R}}\geq 1$.} Nevertheless, such a solution approach fails to offer an interpretation of the common iterative schemes of joint channel / data estimation as outlined in Remark~2) below. 
On the other hand, assuming (as in~\cite{js03}) non-perfect frequency synchronization, involving nonzero CFOs (probably different per antenna~\cite{js03}), the corresponding factor matrix can be assumed to be of full k-rank, which leads to the generic condition
\begin{equation}
M+\min(N,MN_{\mathrm{T}})+\min(N_{\mathrm{R}},MN_{\mathrm{T}})\geq 2MN_{\mathrm{T}}+2
\label{eq:MIMOcondition}
\end{equation}
In practice, where $N$ would probably be larger than $MN_{\mathrm{T}}$, this simplifies to
\[
N_{\mathrm{R}}\geq M(N_{\mathrm{T}}-1)+2
\] 
For the SIMO scenario, this becomes $N_{\mathrm{R}}\geq 2$, which simply requires the spatial dimension to be nontrivial.
Using appropriate precoding at the transmitter can result in more flexible identifiability conditions (e.g., not requiring an excessively large number of receive antennas) and algorithms; see, e.g., \cite{js03,lcsa13}.

\subsection{An ALS view of the joint channel estimation / data detection procedure}

The problem in~\pref{eq:prob} can then be solved with the aid of the classical alternating least squares (ALS) algorithm, iteratively alternating between the  conditional updates (cf.~\pref{eq:Y2eq}, \pref{eq:Y3eq})
\begin{equation}
\g{C}=(\g{H} \odot \g{\Gamma})^{\dagger}\gss{Y}{2}{}
\label{eq:Citer}
\end{equation}
and 
\begin{equation}
\g{H}=\left[(\gss{C}{}{\mathrm{T}}\odot\g{\Gamma})^{\dagger}\gss{Y}{3}{}\right]^{\mathrm{T}},
\label{eq:Hiter} 
\end{equation}
where $(\cdot)^{\dagger}$ stands for the (left) pseudo-inverse. A necessary condition for its existence in~\pref{eq:Hiter} is that there are at least as many Rx as Tx antennas. 
Observe that the permutation ambiguity is trivially resolved in this context because one of the factor matrices, $\g{\Gamma}$, is known, similarly with~\cite{bs02}. A straightforward (and common) way to address the scaling ambiguity is through the transmission of a short training preamble.\footnote{Alternatively ways include appropriate normalization of one of the factors (e.g.,~\cite{rkx12}) or the transmission of a pilot sequence at one of the subcarriers (e.g., \cite{js03,bs02}).} At convergence, and once the complex scaling ambiguity has been resolved, the transmitted symbols can be detected as 
\begin{equation}
\gss{D}{}{(t)}=\mathrm{dec}(\Re\{\gss{C}{}{(t)}\}), \mbox{\ \ } t=1,2,\ldots,N_{\mathrm{T}},
\label{eq:Dt}
\end{equation}
where $\mathrm{dec}(\cdot)$ signifies the decision device for the input constellation. However, this procedure does not exploit the information about $\g{D}$ found in the imaginary (interference) part of~\pref{eq:C} and can be seen to perform similarly or even somewhat worse than CP-OFDM~\cite{kca16}. A significant performance gain can be achieved by taking advantage of the structure of the intrinsic interference through the inclusion of the steps~\pref{eq:Dt} and~\pref{eq:C} between~\pref{eq:Citer} and~\pref{eq:Hiter} in each iteration. Of course, this can only be applied after a few simple iterations of~\pref{eq:Citer}, \pref{eq:Hiter}, due to the presence of the complex scaling ambiguity.

Another important difference with the corresponding OFDM problem is that the noise at the AFB output is colored and hence the cost function in~\pref{eq:prob} should be modified accordingly to a \emph{weighted} LS one. Indeed the noise tensor is correlated in two of its three dimensions (time and frequency, not space) with corresponding covariances that can be \emph{a-priori} known and only depend on the constants $\beta,\gamma,\delta$ (see~\cite{k15} for details). Thus, appropriately modified ALS algorithms can be employed instead (see, e.g.,~\cite{bss02,s04,v05,rkx12}). The reader is referred to the appendix for a more detailed treatment of this subject. 


\noindent
\emph{Remarks.} 
\begin{enumerate}
\item The fact that one of the three factor matrices is known could justify a characterization of the above problem as a bilinear instead of a trilinear one.
To make this explicit, such an ALS algorithm has also been known with the name \emph{bilinear ALS (BALS)}~\cite{rkx12}.
\item One can check that, in a SIMO system, \pref{eq:Citer} and~\pref{eq:Hiter} are in fact nothing but a compact way of re-writing the well-known equations for channel equalization, $c_{p,q}=\frac{1}{N_{\mathrm{R}}}\sum_{r=1}^{N_{\mathrm{R}}}\frac{y^{(r)}_{p,q}}{H_{p}^{(r,1)}}$, and estimation, $H_{p}^{(r,1)}=\frac{1}{N}\sum_{q=0}^{N-1}\frac{y^{(r)}_{p,q}}{c_{p,q}}$.\footnote{The latter is known as Interference Approximation Method (IAM)~\cite{kkrt13}. In fact, as it can be seen in more detail from eqs.~\pref{eq:Hiter_wls} and~\pref{eq:c}, the ALS iterations are equivalent to \emph{maximum-ratio combining (MRC)} operations~\cite{kw97}, which may be simplified (e.g., when all (virtual) symbols have the same magnitude) to the above (IAM) expressions (equivalent to \emph{equal-gain combining (ECG)}~\cite{kw97}.}
It is also of interest to note the similarity of the above ALS procedure with the iterative block algorithms studied in~\cite{tvp96} for the solution of the (\emph{bilinear}) blind maximum likelihood source separation problem, especially the so-called \emph{iterative least squares with projection (ILSP)} scheme. Similarly with~\cite{tvp96}, one could also consider replacing~\pref{eq:Citer} and~\pref{eq:Dt} by a step of enumeration over the input constellation, giving rise to an \emph{iterative least squares with enumeration (ILSE)} scheme, shown in~\cite{tp97} to be generally more effective than the ILSP one. The subsequent increase of complexity could be addressed with the aid of an efficient lattice search (e.g., sphere decoding~\cite{rkpt10}) procedure. 
\item In the present context, the identifiability (uniqueness) question should also consider the discrete (in fact, finite) nature of the set of possible values of the $\g{C}$ factor. No such uniqueness results are known to exist for general 3-way tensors. Nevertheless, one could consider using arguments analogous to those followed in~\cite{tvp96} to show that identifiability is ensured for large enough sets of i.i.d. input symbols. Indeed, since the factors in~\pref{eq:Y2eq} are both of full rank, the identifiability condition of~\cite{tvp96} applies, whereby it suffices for $N$ to be large enough so that $\g{C}$ contains all $\frac{Q^{2M}}{2}$ distinct (up to a sign) $M$-vectors with entries belonging to the $Q^2$-QAM constellation. The probability of non-identifiability for $N\gg \frac{Q^{2M}}{2}$ i.i.d. (multicarrier) symbols is shown in~\cite{tvp96} to approach zero exponentially fast. For large $M$ and/or $Q$, the number of symbols required may become unrealistically large. More practical conditions can be found in, e.g., \cite{lpv03}, albeit only for constant modulus (e.g., QPSK) signals.\footnote{Thanks to Dr.~M.\ S\o rensen, KULeuven, for pointing out this paper.} See also \cite{js03} for a related upper bound on the probability of non-identifiability for the case of i.i.d. BPSK input. It must be noted, of course, that no such problem was encountered in the simulations run for this work. Moreover, as confirmed in the example of the next section, the imaginary part of $\g{C}$ in FBMC/OQAM is close to be Gaussian distributed, providing an extra support to the use of generic rank results that are known to hold for matrices generated by absolutely continuous distributions~\cite{js03}.  
\end{enumerate}

\section{Numerical Examples}

\begin{figure}
\centering
\includegraphics[width=7.5cm]{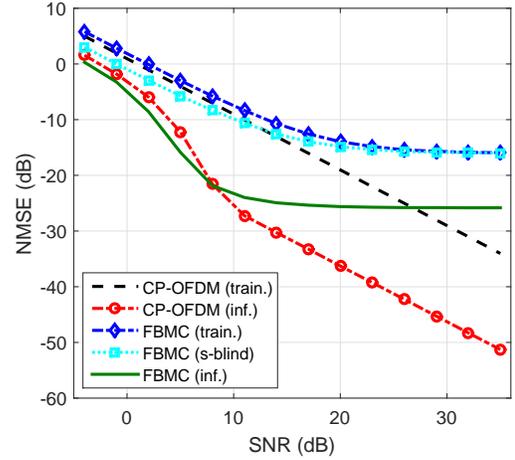} \\
(a) 
\\
\includegraphics[width=7.5cm]{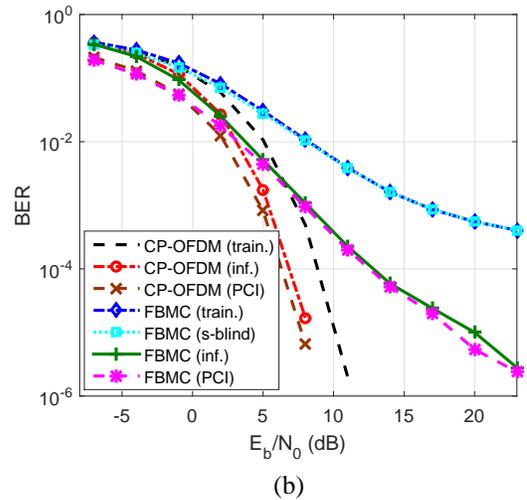} \\
 (b) 
\\
\includegraphics[width=7.5cm]{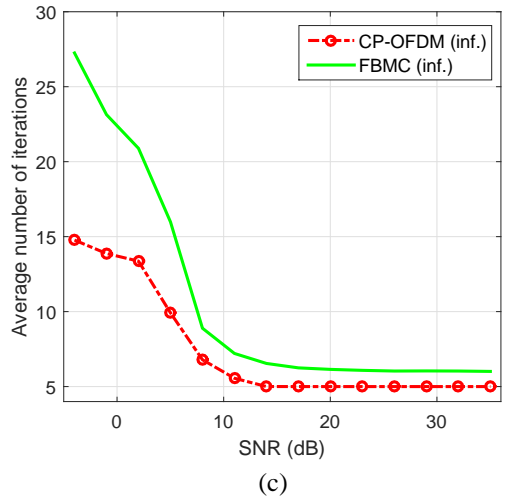} \\
 (c) 
  \caption{Performance comparison for a $1\times 2$ system with PedA channels: (a) NMSE (b) BER (c) average number of iterations.}
  \label{fig:PedA}
\end{figure}

\begin{figure}
\centering
\includegraphics[width=7.5cm]{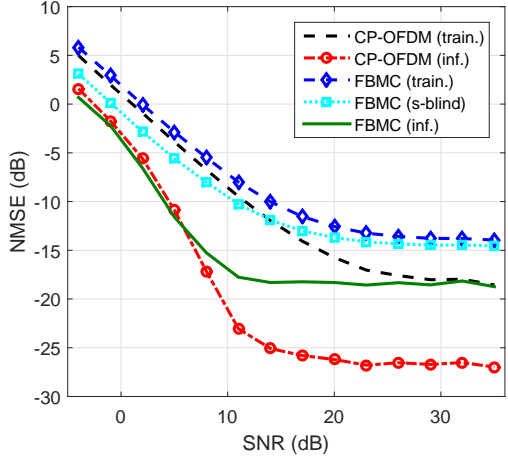} 
  \caption{As in Fig.~\ref{fig:PedA}(a) with VehB channels.} 
  \label{fig:VehB}
\end{figure}

\begin{figure}
\centering
\includegraphics[width=7.5cm]{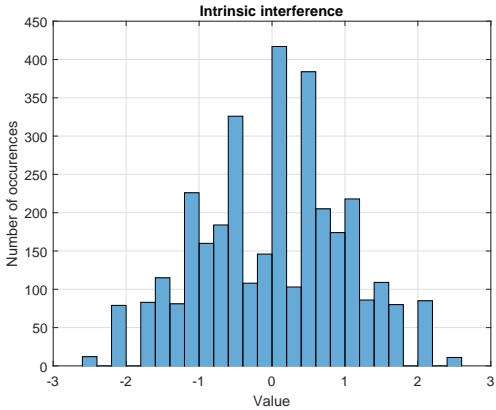} 
  \caption{Histogram of the imaginary part of the pseudo-symbols.} 
  \label{fig:interf}
\end{figure}

The above approach is evaluated here in a SIMO $2\times 1$ system. The input signal is organized in frames of 53~OFDM (i.e., $N=106$~FBMC) symbols each, using QPSK modulation. Filter banks designed as in~\cite{b01} are employed, with $M=32$ and $K=4$. With a subcarrier spacing of 15~kHz, the (block fading) PedA channels involved are of length $L_h=9$ and satisfy the model assumption~\pref{eq:y=Hc+w} only very crudely. The results are compared with those for a similar SIMO-OFDM system, using a CP of $\frac{M}{4}=8$ samples. The estimation performance, in terms of normalized mean squared error (NMSE) versus signal to noise ratio (SNR), is shown in Fig.~\ref{fig:PedA}(a).
In the proposed approach, the \emph{inf}ormed (of the interference structure and input constellation) iterations start after only a couple of initial simple iteration that determines the correct scaling (``inf." curves in the figure). The initialization is random (except, of course, for the known preamble of one OFDM symbol used to deal with the scaling ambiguity problem). It is clearly seen that taking the interference structure and the data constellation into account results in considerable performance gains over the structure-blind approach (see ``s-blind" curves) and outperforms CP-OFDM at low to medium SNR values (at the cost of a larger number of iterations -- albeit of a large variance at low SNR -- as shown in Fig.~\ref{fig:PedA}(c)). Moreover, as expected, jointly estimating the channel and the data symbols brings significant improvement over the training only-based (non data-aided) approach (``train." curves). Analogous conclusions can be drawn from the bit error rate (BER) detection performance depicted in Fig.~\ref{fig:PedA}(b). Notably, the informed approach is observed to yield results quite close to those obtained when perfect channel information (``PCI") is available. The FBMC curves are seen to \emph{floor} at higher SNR values, resulting in performance losses compared to CP-OFDM at such SNR regimes. This is a typical effect of the residual intrinsic interference which comes from the invalidation of model~\pref{eq:y=Hc+w} and shows up in the absence of strong noise~\cite{kkrt13}. These error floors are more severe for more frequency selective channels~\cite{k14}, as shown in the example of VehB channels of length $L_h=18$ depicted in Fig.~\ref{fig:VehB}. 
The CP-OFDM error floors in that case are due to the inadequately long CP used. A rough estimate of the probability density of the intrinsic interference (i.e., the imaginary part of $\g{C}$) is depicted in Fig.~\ref{fig:interf}.  
As previously observed~\cite{PHYDYAS_D3.1}, it resembles a Gaussian density with zero mean (and would look more Gaussian if interference from a wider time-frequency neighborhood or a higher-order input constellation had been considered).

\section{Conclusions}
\label{sec:concls}

The problem of joint channel estimation / data detection in a MIMO-FBMC/OQAM system based on limited training input was studied in this paper from a tensor decomposition-based point of view. Similarly with earlier related MIMO-OFDM works, a 3-way CPD tensor model was shown to provide an accurate description of the system. This resulted in an ALS-type algorithm for determining the channel and symbol matrices, which, in the SIMO setup, reduces to the well-known iterative procedure of jointly estimating the channel and the virtual Tx symbols. A distinctive characteristic of the algorithm, which was demonstrated to offer significant performance gains, is that it takes the constellation of the input and the structure of the intrinsic interference into account. Superior performance in terms of both estimation and detection accuracy over the non-structure aided and the training only-based approaches was demonstrated via simulation results for both mildly and highly frequency selective channels.

On-going work aims at taking  
non-perfect synchronization also into account. 
Further extensions to this work will include channels of strong frequency- (not satisfying~\pref{eq:y=Hc+w}) and time- (e.g., \cite{afm08,mgh08,crkh14}) selectivity, as well as richer configurations involving precoders and space-time/frequency coding (e.g., \cite{lcsa13}).

\appendix

\section*{Colored Noise Considerations}

The discussion in this appendix focuses on the SIMO setup.
Consider first~\pref{eq:Y3eq} in its equivalent vectorized form,
\begin{equation}
\underbrace{\ve(\gss{Y}{3}{})}_{\gss{y}{3}{}}=\left[\gss{I}{N_{\mathrm{R}}}{}\otimes (\gss{C}{}{\T}\odot \boldsymbol{\Gamma})\right]
\underbrace{\ve(\gss{H}{}{\T})}_{\g{h}}+\underbrace{\ve(\gss{W}{3}{})}_{\gss{w}{3}{}},
\label{eq:Y3vec}
\end{equation}
where use has been made of the well-known property of the vectorized form of a matrix product~\cite{sdfhpf16}.  
In view of the assumption that the noise signals at the receiver front ends are identically distributed (with zero mean) and uncorrelated to each other, the covariance matrix of the noise vector  
\begin{eqnarray}
\lefteqn{\gss{w}{3}{}} & & \!\!\!\!\!\! \label{eq:w3} \\
& = & \!\!\!\!\!\! \bmx{cccc} \ve(\gss{W}{}{(1)})^{\T} & \ve(\gss{W}{}{(2)})^{\T} & \cdots & \ve(\gss{W}{}{(N_{\mathrm{R}})})^{\T} \emx^{\T} \nonumber 
\end{eqnarray}
will be of the form
\begin{equation}
\gss{C}{\sbm{w}_3}{}=E\{\gss{w}{3}{}\gss{w}{3}{\He}\}=\gss{I}{N_{\mathrm{R}}}{}\otimes \gss{\bar{C}}{\sbm{w}_3}{},
\label{eq:Cw3}
\end{equation}
where the $MN\times MN$ matrix $\gss{\bar{C}}{\sbm{w}_3}{}$ has the following block tridiagonal, block Toeplitz structure
\begin{eqnarray}
\gss{\bar{C}}{\sbm{w}_3}{} \!\!\!\!\! & = & \!\!\!\!\! \sigma^{2}
\bmx{cccccc} 
\g{B} & \g{S}\gss{A}{}{+} & \g{0} & \g{0} & \cdots & \g{0} \\ \g{S}\gss{A}{}{-} & \g{B} & \g{S}\gss{A}{}{+} & \g{0} & \cdots & \g{0} \\
\g{0} & \g{S}\gss{A}{}{-} & \g{B} & \g{S}\gss{A}{}{+} & \cdots & \g{0} \\
\ddots & \ddots & \ddots & \ddots & \ddots & \ddots \\
\g{0} & \cdots & \g{0} & \g{S}\gss{A}{}{-} & \g{B} & \g{S}\gss{A}{}{+} \\
\g{0} & \cdots & \g{0} & \g{0} & \g{S}\gss{A}{}{-} & \g{B}
\emx \nonumber \\
\!\!\!\!\! & \equiv & \!\!\!\!\! \sigma^{2}\g{\bar{B}},
\label{eq:Cbarw3}
\end{eqnarray}
with the $M\times M$ matrix 
\begin{equation}
\g{B}=\bmx{cccccc} 1 & \jim\beta & 0 & \cdots & 0 & \jim\beta \\ -\jim\beta & 1 & \jim\beta & \cdots & 0 & 0 \\ \vdots & \ddots & \ddots & \ddots & \ddots & \vdots \\ -\jim\beta & 0 & 0 & \cdots & -\jim\beta & 1 \emx
\label{eq:B}
\end{equation}
being the (normalized) covariance matrix of the FBMC noise in the frequency direction, and the $M\times M$ blocks $\g{S}\gss{A}{}{+}$ and $\g{S}\gss{A}{}{-}=(\g{S}\gss{A}{}{+})^{\He}$ standing for the noise correlation in the positive and negative time directions, respectively~\cite{k15}. The $M\times M$ matrices $\g{S}$ and $\gss{A}{}{\pm}$ are given by~\cite{k15}
\begin{eqnarray}
\g{S} \!\!\!\!\! & = & \!\!\!\!\! \diag(1,-1,1,-1,\ldots,1,-1), \label{eq:S} \\
\gss{A}{}{\pm} \!\!\!\!\! & = & \!\!\!\!\! \jim 
\bmx{cccccccc} 
\pm\gamma & \delta & 0 & 0 & 0 & \cdots & 0 & -\delta \\ 
\delta & \pm\gamma & \delta & 0 & 0 & \cdots & 0 & 0 \\
0 & \delta & \pm\gamma & \delta & 0 & \cdots & 0 & 0 \\
\vdots & \ddots & \ddots & \ddots & \ddots & \vdots & \ddots & \ddots \\
-\delta & 0 & 0 & 0 & 0 & \cdots & \delta & \pm\gamma 
\emx
\label{eq:Apm}
\end{eqnarray}
The block diagonal structure of~\pref{eq:Cw3} (due to the fact that the noise is uncorrelated in the spatial dimension) implies that the channel responses corresponding to the $N_{\mathrm{R}}$ receive antennas can be estimated separately. To see this in detail, consider the maximum likelihood (ML) estimate as resulting  from~\pref{eq:Y3vec}~\cite{k93}
\begin{eqnarray}
\g{\hat{h}} & = & \bmx{cccc} (\gss{H}{}{(1)})^{\T} & (\gss{H}{}{(2)})^{\T} & \cdots & (\gss{H}{}{(N_{\mathrm{R}})})^{\T} \emx^{\T} \nonumber \\
& = & \left[(\gss{I}{N_{\mathrm{R}}}{}\otimes (\gss{C}{}{\T}\odot \boldsymbol{\Gamma}))^{\He}\gss{C}{\sbm{w}_3}{-1}(\gss{I}{N_{\mathrm{R}}}{}\otimes (\gss{C}{}{\T}\odot \boldsymbol{\Gamma}))\right]^{-1}\times \nonumber \\
& & (\gss{I}{N_{\mathrm{R}}}{}\otimes (\gss{C}{}{\T}\odot \boldsymbol{\Gamma}))^{\He}\gss{C}{\sbm{w}_3}{-1}\gss{y}{3}{}
\label{eq:h1}
\end{eqnarray}
or equivalently, invoking~\pref{eq:Cw3}, \pref{eq:Cbarw3} and properties of the Kronecker product~\cite{sdfhpf16}, as in~\pref{eq:h}, at the top of the next page.
\begin{figure*}
\begin{equation}
\g{\hat{h}} = \left[\gss{I}{N_{\mathrm{R}}}{}\otimes \left(\left((\gss{C}{}{\T}\odot \boldsymbol{\Gamma})^{\He}\gss{\bar{B}}{}{-1}(\gss{C}{}{\T}\odot \boldsymbol{\Gamma})\right)^{-1}(\gss{C}{}{\T}\odot \boldsymbol{\Gamma})^{\He}\gss{\bar{B}}{}{-1}\right)\right]\gss{y}{3}{}
\label{eq:h}
\end{equation}
\hrulefill
\end{figure*}
This implies that each of the channel estimates can be computed as
\begin{eqnarray}
\lefteqn{\gss{\hat{H}}{}{(r)}=} & & \nonumber \\
& & \!\!\!\!\! \left[(\gss{C}{}{\T}\odot \boldsymbol{\Gamma})^{\He}\gss{\bar{B}}{}{-1}(\gss{C}{}{\T}\odot \boldsymbol{\Gamma})\right]^{-1}(\gss{C}{}{\T}\odot \boldsymbol{\Gamma})^{\He}\gss{\bar{B}}{}{-1}\ve(\gss{Y}{}{(r)}) \nonumber \\
& = & \left[\bmx{c} \diag(\gss{c}{0}{}) \\ \diag(\gss{c}{1}{}) \\ \vdots \\ \diag(\gss{c}{N-1}{}) \emx^{\He}\gss{\bar{B}}{}{-1}
\bmx{c} \diag(\gss{c}{0}{}) \\ \diag(\gss{c}{1}{}) \\ \vdots \\ \diag(\gss{c}{N-1}{}) \emx\right]^{-1} \times \nonumber \\ 
& & \bmx{c} \diag(\gss{c}{0}{}) \\ \diag(\gss{c}{1}{}) \\ \vdots \\ \diag(\gss{c}{N-1}{}) \emx^{\He}\gss{\bar{B}}{}{-1}\ve(\gss{Y}{}{(r)}),
\label{eq:Hiter_wls}
\end{eqnarray}
for $r=1,2,\ldots,N_{\mathrm{R}}$.

In an analogous manner, eq.~\pref{eq:Y2eq} can be written in vectorized form as
\begin{equation}
\underbrace{\ve(\gss{Y}{2}{})}_{\gss{y}{2}{}}=\left[\gss{I}{N}{}\otimes (\g{H}\odot \boldsymbol{\Gamma})\right]\underbrace{\ve(\g{C})}_{\g{c}}
+\underbrace{\ve(\gss{W}{2}{})}_{\gss{w}{2}{}}
\label{eq:Y2vec}
\end{equation}
where
\begin{equation}
\gss{W}{2}{}=\bmx{cccc} (\gss{W}{}{(1)})^{\T} & (\gss{W}{}{(2)})^{\T} & \cdots & (\gss{W}{}{(N_{\mathrm{R}})})^{\T}\emx^{\T}
\label{eq:W2}
\end{equation}
is a re-arrangement of $\gss{W}{3}{}$. In fact, it is readily verified that their vectorized versions are related as
\begin{equation}
\gss{w}{2}{}=\gss{P}{2,3}{}\gss{w}{3}{},
\label{eq:w2=Pw3}
\end{equation}
where $\gss{P}{2,3}{}$ is a permutation matrix of order $N_{\mathrm{R}}NM$, given by
\begin{equation}
\gss{P}{2,3}{}=\gss{I}{N_{\mathrm{R}},N}{}\otimes \gss{I}{M}{},
\label{eq:P}
\end{equation}
with $\gss{I}{N_{\mathrm{R}},N}{}$ denoting (in the notation of~\cite{hs81}) the vec-permutation matrix of order $N_{\mathrm{R}}N$.
It then follows that the covariance matrix of $\gss{w}{2}{}$ is~\cite{v05}
\begin{equation}
\gss{C}{\sbm{w}_2}{}=\gss{P}{2,3}{}\gss{C}{\sbm{w}_3}{}\gss{P}{2,3}{\T}
\label{eq:Cw2=PCw3P}
\end{equation}
Using the previous relations in the analogous to~\pref{eq:h1} expression for the ML estimate of $\g{c}$ yields
\begin{eqnarray}
\lefteqn{\g{\hat{c}}=} & & \nonumber \\
& & \left[(\gss{I}{N}{}\otimes (\g{H}\odot \boldsymbol{\Gamma})^{\He})\gss{P}{2,3}{}\gss{C}{\sbm{w}_3}{-1}\gss{P}{2,3}{\T}
(\gss{I}{N}{}\otimes (\g{H}\odot \boldsymbol{\Gamma}))\right]^{-1} \times \nonumber \\
& & (\gss{I}{N}{}\otimes (\g{H}\odot \boldsymbol{\Gamma})^{\He})\gss{P}{2,3}{}\gss{C}{\sbm{w}_3}{-1}\gss{P}{2,3}{\T}\gss{y}{2}{}
\label{eq:c1}
\end{eqnarray}
Note that, similarly with~\pref{eq:w2=Pw3},
\begin{equation}
\gss{P}{2,3}{\T}\gss{y}{2}{}=\gss{y}{3}{}
\label{eq:y2=Py3}
\end{equation}
Furthermore, it is not difficult to see that
\begin{equation}
\gss{P}{2,3}{\T}(\gss{I}{N}{}\otimes (\g{H}\odot \boldsymbol{\Gamma}))=\bmx{c} \gss{I}{N}{}\otimes \diag(\gss{H}{}{(1)}) \\ \vdots \\ 
\gss{I}{N}{}\otimes \diag(\gss{H}{}{(N_{\mathrm{R}})}) \emx
\label{eq:P[.]}
\end{equation}
Invoking these relations in~\pref{eq:c1} together with~\pref{eq:Cw3} leads to the more explicit expression for the estimate of the symbols vector shown at the top of the next page (eq.~\pref{eq:c}).
\begin{figure*}
\begin{equation}
\g{\hat{c}}=
\left[\sum_{r=1}^{N_{\mathrm{R}}}(\gss{I}{N}{}\otimes \diag(\gss{H}{}{(r)*}))\gss{\bar{B}}{}{-1}(\gss{I}{N}{}\otimes \diag(\gss{H}{}{(r)}))\right]^{-1} 
\sum_{r=1}^{N_{\mathrm{R}}}(\gss{I}{N}{}\otimes \diag(\gss{H}{}{(r)*}))\gss{\bar{B}}{}{-1}\ve(\gss{Y}{}{(r)})
\label{eq:c}
\end{equation}
\hrulefill
\end{figure*}

For the sake of completeness, consider also the covariance of $\gss{w}{1}{}=\ve(\gss{W}{1}{})$ in~\pref{eq:Y1eq}. It is easily verified that 
the permutation matrix transforming $\gss{w}{3}{}$ to $\gss{w}{1}{}$ is given by
\begin{equation}
\gss{P}{1,3}{}=\gss{I}{N_{\mathrm{R}}N,M}{},
\label{eq:P13}
\end{equation}
hence
\begin{equation}
\gss{C}{\sbm{w}_1}{}=\gss{P}{1,3}{}\gss{C}{\sbm{w}_3}{}\gss{P}{1,3}{\T}
\label{eq:Cw1}
\end{equation}

It must be emphasized here that, despite the rich structure of the matrix $\g{\bar{B}}$, namely block tridiagonal\footnote{More generally, block banded (if the prototype filter and the overlapping factor are such that there exists non-negligible correlation in time beyond the immediate neighbors).}, block Toeplitz, Hermitian, with structured (banded circulant) blocks, its inversion is far from being an easy problem (e.g., \cite{gcb07,bhr15}) and asks for specialized algorithmic solutions, particularly in view of its large scale in practice (e.g., \cite{m13}). Notably, the square-root factorization of $\g{\bar{B}}$ for the small case of $N=2$ was recently developed in~\cite{k15} and proved to be far from being straightforward. In view of the above, and for the sake of simplicity, the noise correlation due to the FBMC modulation is neglected in this paper. Note that, with $\g{\bar{B}}$ set to the identity matrix, eqs.~\pref{eq:Hiter_wls} and~\pref{eq:c} also describe the ALS iterations for the OFDM system.
 
%
%
%



\begin{thebibliography}{99} 
\bibitem{a07} A.\ L.\ F.\ de Almeida, \emph{Tensor Modeling and Signal Processing for Wireless Communication Systems}, PhD thesis, University of Nice (Sophia Antipolis), 2007. 
\bibitem{afm06} A.\ L.\ F.\ de Almeida, G.\ Favier, J.\ C.\ M.\ Mota, and R.\ L. de Lacerda, ``Estimation of frequency-selective block-fading MIMO channels using PARAFAC modeling and alternating least squares," \emph{Proc.~40th~Asilomar Conf. Signals, Systems, and Computers}, Pacific~Grove, CA, 29~Oct.--1~Nov.~2006.
\bibitem{afm08} A.\ L.\ F.\ de Almeida, J.\ C.\ M.\ Mota, and G.\ Favier, ``PARAFAC modeling/estimation of time-varying space-time multipath channels," 
\emph{Proc.~Congresso Tecnol\'{o}gico TI \& Telecom}, Fortaleza, Brazil, Aug.~2008. 
\bibitem{bn10} Q.\ Bai and J.\ A.\ Nossek, ``On the effects of carrier frequency offset on cyclic prefix based {OFDM} and filter bank based multicarrier systems," \emph{Proc.~SPAWC-2010}, Marrakech, Morocco, June~2010. 
\bibitem{bbcmru14} P.\ Banelli \emph{et al.}, ``Modulation formats and waveforms for 5G networks: Who will be the heir of {OFDM}?: An overview of alternative modulation schemes for improved spectral efficiency," \emph{IEEE Signal Process.\ Mag.}, vol.~31, no.~6, pp.~80--93, Nov.~2014. 
\bibitem{bda03} A.\ de Baynast, L.\ De Lathauwer, and B.\ Aazhang, ``Blind {PARAFAC} receivers for multiple access-multiple antenna systems,"
\emph{Proc.~VTC-2003 (Fall)}, Orlando, FL, Oct.~2003. 
\bibitem{b01} M.\ G.\ Bellanger, ``Specification and design of a prototype filter for filter bank based multicarrier transmission," \emph{Proc.~ICASSP-2001}, Salt Lake City, UT, May~2001. 
\bibitem{bdh01} H.\ B\"{o}lcskei, P.\ Duhamel, and R.\ Hleiss, ``A subspace-based approach to blind channel identification in pulse shaping {OFDM/OQAM} systems,"
\emph{IEEE Trans.\ Signal Process.}, vol.~49, no.~7, pp.~1594--1598, July~2001. 
\bibitem{bhr15} N.\ M.\ Boffi, J.\ C.\ Hill, and M.\ G.\ Reuter, ``Characterizing the inverses of block tridiagonal, block Toeplitz matrices,"
\emph{Computational Science \& Discovery}, vol.~8, 2015.
\bibitem{bhsl09} R.\ Bro, R.\ A.\ Harshman, N.\ D.\ Sidiropoulos, and M.\ E.\ Lundy, ``Modeling multi-way data with linearly dependent loadings," \emph{J.\ Chemometrics}, vol.~23, pp.~324--340, 2009.
\bibitem{bss02} R.\ Bro, N.\ D.\ Sidiropoulos, and A.\ K.\ Smilde, ``Maximum likelihood fitting using ordinary least squares algorithms," \emph{J.\ Chemometrics},
vol.~16, pp.~387--400, 2002. 
\bibitem{bs02} R.\ S.\ Budampati and N.\ D.\ Sidiropoulos, ``Khatri-Rao space time codes with maximum diversity gains over frequency-selective channels," \emph{Proc.~SAM-2002}, Rosslyn, VA, Aug.~2002. 
\bibitem{crkh14} Y.\ Cheng, F.\ Roemer, O.\ Khatib, and M.\ Haardt, ``Tensor subspace Tracking via Kronecker structured projections (TeTraKron) for time-varying multidimensional harmonic retrieval," \emph{EURASIP J. Advances in Signal Process.}, 2014 (http://asp.eurasipjournals.com/content/2014/1/123). 
\bibitem{dd13} I.\ Domanov and L.\ De\ Lathauwer, ``On the uniqueness of the Canonical Polyadic Decomposition of third-order tensors -- Part ~II: Uniqueness of the overall decomposition," \emph{SIAM Journal on Matrix Analysis and Applications},  pp.~855--875, vol.~34, no.~3, 2013.
\bibitem{f11} B.\ Farhang-Boroujeny, ``{OFDM} vs.\ filter bank multicarrier," \emph{IEEE Signal Process.\ Mag.}, vol.~28, no.~3, pp.~92--112, May~2011.
\bibitem{fa14} G.\ Favier and A.\ L. F.\ de Almeida, ``Overview of constrained PARAFAC models," \emph{EURASIP Journal on Advances in Signal Processing}, vol.~2014, no.~142, 2014. 
\bibitem{fzwlx11} G.\ Feng, X.\ Zhang, H.\ Wu, J.\ Li, and D.\ Xu, ``Novel carrier frequency offset estimator in {MIMO-OFDM} system," \emph{International Journal of Digital Content Technology and its Applications}, vol.~5, no.~3, pp.~59--66, March~2011.
\bibitem{ghst01} G.\ B.\ Giannakis, Y.\ Hua, P.\ Stoica, and L.\ Tong (eds.), \emph{Signal Processing Advances in Wireless and Mobile Communications},
Prentice-Hall, 2001.
\bibitem{gcb07} J.\ Guti\'{e}rrez-Guti\'{e}rrez, P.\ M.\ Crespo, and A.\ B\"{o}ttcher, ``Functions of banded Hermitian block Toeplitz matrices
in signal processing," \emph{Linear Algebra and its Applications}, vol.~422, pp.~788–-807, 2007. 
\bibitem{gl13} G.\ Gong and J.\ Liu, ``MIMO-OFDM system estimation via three-dimensional channel decomposition," \emph{Proc.~ICMT-2013}, Guangzhou, China,
29~Nov.--1st~Dec.~2013. 
\bibitem{hs81} H.\ V.\ Henderson and S.\ R.\ Searle, ``The vec permutation matrix, the vec operator and Kronecker products: A review," \emph{Linear and Multilinear Algebra}, vol.~9, no.~4, pp.~271--288, 1981. 
\bibitem{hc14} W.\ Hou and B.\ Champagne, ``Semiblind channel estimation for {OFDM/OQAM} systems," \emph{IEEE Signal Process.\ Letters}, vol.~22, no.~4, pp.~400--403, Oct.~2014. 
\bibitem{js03} T.\ Jiang and N.\ D.\ Sidiropoulos, ``A direct blind receiver for SIMO and MIMO OFDM systems subject to unknown offset and multipath,"
\emph{Proc.~SPAWC-2003}, Rome, Italy, June~2003. 
\bibitem{js04} T.\ Jiang and N.\ D.\ Sidiropoulos, ``Blind identification of out of cell users in DS-CDMA," \emph{EURASIP Journal on Applied Signal Processing}, pp.~1212--1224, Aug.~2004.
\bibitem{k93} S.\ M.\ Kay, \emph{Fundamentals of Statistical Signal Processing -- Vol.~I: Estimation Theory}, Prentice-Hall, 1993.
\bibitem{k14} E.\ Kofidis, ``Channel estimation in filter bank-based multicarrier systems: Challenges and solutions," \emph{Proc.~ISCCSP-2014}, Athens, Greece, May~2014. 
\bibitem{k15} E.\ Kofidis, ``On optimal multi-symbol preambles for highly frequency selective {FBMC/OQAM} channel estimation," \emph{Proc.~ISWCS-2015}, Brussels, Belgium, Aug.~2015.
\bibitem{kca16} E.\ Kofidis, C.\ Chatzichristos, and A.\ L.\ F.\ de Almeida, ``Tensor-based processing of filter bank-based multicarrier signals," poster presented at the \emph{Workshop on Tensor Decompositions and Applications (TDA-2016)}, Leuven, Belgium, Jan.~2016.
\bibitem{kkrt13} E.\ Kofidis, D.\ Katselis, A.\ Rontogiannis, and S.\ Theodoridis, ``Preamble-based channel estimation in {OFDM/OQAM} systems: A review," \emph{Signal Process.}, vol.~93, pp.~2038--2054, 2013.
\bibitem{kb09} T.\ G.\ Kolda and B.\ W.\ Bader, ``Tensor decompositions and applications," \emph{SIAM Review}, vol.~51, no.~3, pp.~455–-500, 2009.
\bibitem{kw97} R.\ Krenz and K.\ Weso{\l}owski, ``Comparative study of space-diversity techniques for MLSE receivers in mobile radio,"
\emph{IEEE Trans.\ Veh.\ Techn.}, vol.~46, no.~3, pp.~653--663, Aug.~1997. 
\bibitem{lpv03} A.\ Leshem, N.\ Petrochilos, and A.-J.\ van der Veen, ``Finite sample identifiability of multiple constant modulus sources,"
\emph{IEEE Trans.\ Info.\ Theory}, vol.~49, no.~9, pp.~2314--2319, Sept.~2003. 
\bibitem{lcas12} K.\ Liu, J.\ P.\ C.\ L.\ da Costa, A.\ L.\ F.\ de Almeida, and H.\ C.\ So, 
``A closed form solution to semi-blind joint symbol and channel estimation in {MIMO-OFDM} systems", \emph{Proc.~ICSPCC-2012}, Hong Kong, Aug.~2012. 
\bibitem{lcsa13} K.\ Liu, J.\ P.\ C.\ L.\ da Costa, H.\ C.\ So, and A.\ L.\ F.\ de Almeida, 
``Semi-blind receivers for joint symbol and channel
estimation in space-time-frequency {MIMO-OFDM} systems," \emph{IEEE Trans.\ Signal Process.}, vol.~61, no.~21, pp.~5444--5457, 1~Nov.~2013.
\bibitem{PHYDYAS_D3.1} J.\ Louveaux \emph{et al.}, ``Equalization and demodulation in the receiver (single antenna)," 
Deliverable~D3.1, PHYDYAS ICT project, July~2008 (http://www.ict-phydyas.org/delivrables/PHYDYAS-D3.1.pdf/view).  
\bibitem{lgss13} J.\ A.\ L\'{o}pez-Salcebo, E.\ Gutierez, G.\ Seco-Granados, and A.\ Lee~Swindlehurst, ``Unified framework for the synchronization of flexible multicarrier communication signals," 
\emph{IEEE Trans.\ Signal Process.}, vol.~61, no.~4, pp.~828--842, Feb.~2013.
\bibitem{m13} A.\ N.\ Malyshev, ``On solution of large systems of linear equations with block-Toeplitz banded matrices," \emph{Doklady Mathematics}, vol.~87, no.~2, pp.~153–-155, 2013.
\bibitem{mt16} D.\ Mattera and M.\ Tanda, ``Optimum single-tap per-subcarrier equalization for {OFDM/OQAM} systems," \emph{Digital Signal Processing}, vol.~49, pp.~148--161, Feb.~2016.
\bibitem{mgh08} M.\ Milojevi\'{c}, G.\ Del Galdo, and M.\ Haardt, ``Tensor-based framework for the prediction of frequency-selective time-variant MIMO channels," 
\emph{Proc.~WSA-2008}, Vienna, Austria, Feb.~2008. 
\bibitem{mfmfcf15} J.\ P.\ Miranda, A.\ Farhang, N.\ Marchetti, F.\ A.\ P.\ de Figueiredo, F.\ A.\ C.\ M.\ Cardoso, and F.\ Figueiredo, ``On massive {MIMO} and its applications to machine type communications and {FBMC}-based networks,"
\emph{EAI Endorsed Trans.\ Ubiquitous Environments}, vol.~15, no.~5, July~2015. 
\bibitem{nl07} D.\ Nion and L.\ De Lathauwer, ``Blind receivers based on tensor decompositions -- Application in DS-CDMA and over-sampled systems," \emph{Proc.~41st Asilomar Conf.\ Signals, Systems, and Computers}, Pacific~Grove, CA, Nov.~2007.
\bibitem{rcs06} M.\ Rajih, P.\ Comon, and D.\ Slock, ``A deterministic blind receiver for {MIMO-OFDM} systems," \emph{Proc.~SPAWC-2006},
Cannes, France, July~2006. 
\bibitem{r15} M.\ Rezaee, ``Low complexity closed-form solution to blind joint channel and symbol estimation in {MIMO-OFDM}," 
\emph{International Journal of Electrical \& Computer Sciences}, vol.~15, no.~3, pp.~1--6, June~2015. 
\bibitem{rrc15} M.\ Rezaee, S.\ R.\ F.\ Rosa, and V.\ C.\ G.\ Coelho, ``A closed-form semi-blind joint channel and symbol estimation approach for {MIMO-OFDM}," 
\emph{International Journal of Science and Advanced Technology}, vol.~5, no.~6, pp.~41--46, June~2015. 
\bibitem{rkpt10} C.\ Rizogiannis, E.\ Kofidis, C.\ B.\ Papadias, and S.\ Theodoridis, ``Semi-blind maximum-likelihood joint channel/data estimation for correlated channels in multiuser MIMO networks," \emph{Signal Processing}, vol.~90, no.~4, pp.~1209--1224, April~2010. 
\bibitem{rh10} F.\ Roemer and M.\ Haardt, ``Tensor-based channel estimation and iterative refinements for two-way relaying with multiple antennas and spatial reuse," \emph{IEEE Trans.\ Signal Process.}, vol.~58, no.~11, pp.~5720--5735, Nov.~2010. 
\bibitem{rkx12} Y.\ Rong, M.\ R.\ A.\ Khandaker, and Y.\ Xiang, ``Channel estimation of dual-hop MIMO relay system via parallel factor analysis,"
\emph{IEEE Trans. Wireless Commun.}, vol.~11, no.~6, pp.~2224--2233, June~2012.
\bibitem{sh10} A.\ Santra and K.\ V.\ S.\ Hari, ``Low complexity {PARAFAC} receiver for {MIMO-OFDMA} system in the presence of multi-access interference,"
\emph{Proc.~44th~Asilomar Conf. Signals, Systems, and Computers}, Pacific~Grove, CA, Nov.~2010. 
\bibitem{sbp16} V.\ Savaux, F.\ Bader, and J. Palicot, ``{OFDM/OQAM} blind equalization using {CNA} approach," \emph{IEEE Trans.\ Signal Process.}, vol.~64, no.~9, 
pp.~2324--2333, May~2016. 
\bibitem{s04} N.\ D.\ Sidiropoulos, ``Low-rank decomposition of multi-way arrays: A signal processing perspective," \emph{Proc.~SAM-2004}, Barcelona, Spain, July~2004.
\bibitem{sb01} N.\ Sidiropoulos and R.\ Bro, ``{PARAFAC} techniques for signal separation," Chapter~4 in~\cite[Vol.~2]{ghst01}.
\bibitem{sb02} N.\ D.\ Sidiropoulos and R.\ S.\ Budampati, ``Khatri-Rao space-time codes," \emph{IEEE Trans.\ Signal Process.}, vol.~50, no.~10, pp.~2396--2407, Oct.~2002. 
\bibitem{sdfhpf16} N.\ D.\ Sidiropoulos, L.\ De~Lathauwer, X.\ Fu, K.\ Huang, E.\ E.\ Papalexakis, and C.\ Faloutsos, ``Tensor decomposition for signal processing and machine learning," \emph{arXiv:1607.01668v1 [stat.ML]}, July~2016. 
\bibitem{sgb00} N.\ D.\ Sidiropoulos, G.\ B.\ Giannakis, and R.\ Bro, ``Blind {PARAFAC} receivers for {DS-CDMA} systems," \emph{IEEE Trans.\ Signal Process.,} vol.~48, no.~3, pp.~810--823, March~2000. 
\bibitem{ssl02} P.\ Siohan, C.\ Siclet, and N.\ Lacaille, ``Analysis and design of {OFDM/OQAM} systems based on filterbank theory,"
\emph{IEEE Trans.\ Signal Process.}, vol.~50, no.~5, pp.~1170--1183, May~2002. 
\bibitem{shr13} B.\ Song, M. Haardt, and F.\ Roemer, ``A tensor-based subspace method for blind estimation of {MIMO} channels," \emph{Proc.~ISWCS-2013}, Ilmenau, Germany, Aug.~2013. 
\bibitem{sd13} M. S\o rensen and L. De Lathauwer, ``Blind signal separation via tensor decomposition with Vandermonde factor: Canonical Polyadic Decomposition,"
\emph{IEEE Trans.\ Signal Process.}, vol.~61, no.~22, pp.~5507--5519, 15~Nov.~2013. 
\bibitem{sd15} M.\ S\o rensen and L.\ De Lathauwer, ``New uniqueness conditions for the canonical polyadic decomposition of third-order tensors," \emph{SIAM Journal on Matrix Analysis and Applications},  vol.~36, no.~4, pp.~381--140, 2015. 
\bibitem{ss07} A.\ Stegeman and N.\ D.\ Sidiropoulos, ``On Kruskal's uniqueness condition for the Candecomp/Parafac decomposition," \emph{Linear Algebra and its Applications}, pp.~540--552, 2007.
\bibitem{s15} B.\ Su, ``Semi-blind channel estimation for OFDM/OQAM systems assisted by zero-valued pilots," \emph{Proc.~DSP-2015}, Singapore, July~2015. 
\bibitem{tvp96} S.\ Talwar, M.\ Viberg, and A.\ Paulraj, ``Blind separation of synchronous co-channel digital signals using an antenna array--Part~I: Algorithms,"
\emph{IEEE Trans.\ Signal Process.}, vol.~44, no.~5, pp.~1184--1197, May~1996.
\bibitem{tp97} S.\ Talwar and A.\ Paulraj, ``Blind separation of synchronous co-channel digital signals using an antenna array--Part~II: Performance analysis,"
\emph{IEEE Trans.\ Signal Process.}, vol.~45, no.~3, pp.~706--718, March~1997.
\bibitem{vsd16} F.\ Van Eeghem, M.\ S\o rensen, and L.\ De~Lathauwer, ``Tensor decompositions with several block-Hankel factors and application in blind system identification," Internal Report 16-39, ESAT-STADIUS, KU Leuven (Leuven, Belgium), 2016.
\bibitem{v05} L.\ J.\ Vega Montoto, \emph{Maximum Likelihood Methods for Three-Way Analysis in Chemistry}, PhD thesis, Dalhousie University, Halifax, Nova Scotia, Canada, April~2005.
\bibitem{xcj12} H.\ Xi, Y.\ Chaowei, and Z.\ Jinbo, ``Blind signal detection algorithm for {OFDMA} uplink systems using {PARALIND} model,"
\emph{International Journal of Digital Content Technology and its Applications}, vol.~6, no.~23, pp.~817--825, Dec.~2012. 
\bibitem{xfd11} Z.\ Xiaofei, W.\ Fei, and X. Dazhuan, ``Blind signal detection algorithm for MIMO-OFDM systems over multipath channel using PARALIND model,"
\emph{IET Commun.}, vol.~5, no.~5, pp.~606--611, 2011. 
\bibitem{zgx10} X.\ Zhang, X.\ Gao, and D.\ Xu, ``Novel blind carrier frequency offset estimation for {OFDM} system with multiple antennas,"
\emph{IEEE Trans. Wireless Commun.}, vol.~9, no.~3, pp.~881--885, March~2010. 
\end{thebibliography}
\end{document}